# Investigating Youths' Everyday Understanding of Machine Learning Applications: a Knowledge-in-Pieces Perspective

Luis Morales-Navarro & Yasmin B. Kafai
luismn@upenn.edu, kafai@upenn.edu
University of Pennsylvania

**Abstract:** Despite recent calls for including artificial intelligence (AI) literacy in K-12 education, not enough attention has been paid to studying youths' everyday knowledge about machine learning (ML). Most research has examined how youths attribute intelligence to AI/ML systems. Other studies have centered on youths' theories and hypotheses about ML highlighting their misconceptions and how these may hinder learning. However, research on conceptual change shows that youths may not have coherent theories about scientific phenomena and instead have knowledge pieces that can be productive for formal learning. We investigate teens' everyday understanding of ML through a knowledge-in-pieces perspective. Our analyses reveal that youths showed some understanding that ML applications learn from training data and that applications recognize patterns in input data and depending on these provide different outputs. We discuss how these findings expand our knowledge base and implications for the design of tools and activities to introduce youths to ML.

## Introduction

The last decade has concentrated on making computing an integral part of K-12 education, with most of these efforts have focused on imperative programming in which people shape and change program states through instructions (Tedre et al., 2021). However, the landscape is evolving with the widespread adoption of artificial intelligence (AI) and machine learning (ML). This paradigm shift requires changes in computing education, as in ML data, rather than code, plays a pivotal role in shaping the behavior of computing systems (for further discussion of this shift see: Tedre et al., 2021; Shapiro & Tissenbaum, 2019). As such, in the past five years, application areas of ML have received increased attention in computing education because of their pervasiveness and societal impact, here it is important for current and future efforts to connect to students' lived experiences.

In their everyday lives, youths already have rich experiences with ML-powered applications (MLPAs), using them while going to school, playing games, listening to music, and socializing with their peers. However, as Long and Magerko (2020) conclude in a recent scoping review, there is not enough empirical evidence to draw "a robust and accurate understanding of what existing pre-conceptions non-programmers have about AI" (p. 10). Another challenge has been how to frame these everyday understandings from a theoretical perspective. Previous research has centered on (a) intelligence attribution studies that explore learners' perspectives on MLPAs but not their explanations on how these work (e.g., Williams et al., 2019), (b) taking a theory approach, as if young people had well-defined coherent theories about how ML works (e.g., Druga & Ko, 2021), and (c) misconceptions in everyday ML knowledge, describing these as obstacles for learning (e.g, Long & Magerko, 2020). Yet research in conceptual change shows that everyday knowledge may not be coherently organized, existing "in-pieces" that include pieces of knowledge that can be productive for learners to understand complex phenomena such as ML (diSessa et al., 2004).

In this study we expand on existing research by applying the theoretical perspective of knowledge-in-pieces (KiP)—extensively used to investigate everyday understandings of physics and mathematics—to the analysis of youths' everyday understandings of MLPAs. We use cooperative inquiry methods to investigate how youths, 14-16 years old, explain how MLPAs work by having them interact with artifacts and explain their functionality. We discuss implications for the design of learning environments and tools and future directions for studying everyday ML understanding.

## Background

Previous research on young people's everyday understanding of MLPAs has adopted different theoretical perspectives based on (1) attributions of intelligence to MLPAs, (2) theories or hypotheses of how applications work, or (3) misconceptions that emerge from these theories. In this paper, we propose a fourth framing, (4) knowledge-in-pieces, to further investigate youths' everyday understanding of ML. In the following sections we introduce each of these approaches and the current study.

### Attributions perspective

Studies from the intelligence attribution perspective center on how young people judge MLPAs in terms of their intelligence and human-like characteristics. Researchers have found that children's perception of their own intelligence when compared to MLPAs depends on different modalities of interaction (Druga et al., 2017). In one study, while most children were unsure of whether MLPAs were smarter than themselves, most children agreed that these "robots" could learn and "always follow the rules" (Williams et al., 2019). In this context, work has investigated the role of parents in supporting how children understand and attribute intelligence to MLPAs (Druga et al., 2018), finding that children's answers were like their parents. Similarly, studies have centered on whether children perceive MLPAs as human-like, with children characterizing some features of applications (such as inaccuracy) as human-like and others (such as lack of common sense) as non-human (Festerling & Siraj, 2020). For example, in a study on youths' perception of Amazon's Alexa, participants reported that Alexa was highly intelligent and safe, moderately trustworthy, and human-like, and not very alive (Van Brummelen et al., 2021). Another study found that perceptions of AI among youths (ages 10-16) centered around robots, reinforcing science fiction ideas present in popular media (Rodríguez-García et al., 2021). Whereas these studies on intelligence attribution and human-like traits provide insights into children's perceptions of ML technologies, do not account for youths' understanding of the inner workings of these technologies.

### Theories perspective

Far less attention has been given to youths' everyday understanding of *how* MLPAs work. Szczuka and colleagues (2022) found that children drew inferences on the functionality of MLPAs (in this study voice assistants) particularly with regards to data processing. Likewise, Druga and Ko (2021) investigated children's hypotheses about the behavior of "smart" artifacts. They argued that children used their funds of knowledge or prior experiences (from interacting with MLPAs and computers to information heard in the media and at home) to create theories about system behaviors and build hypotheses that included egocentric speculations (based on how children themselves would behave) and observational ad-hoc hypotheses (based on the behavior of the agents). While these studies contribute detailed evidence on children's understanding of ML, they seem to align with research in conceptual development that argues that children organize everyday understandings to build consistent hypotheses or theories (Vosniadou, 2019). In other fields such as physics, this hypothesis/theory approach to everyday understandings has led to the belief that everyday theories that differ from scientific explanations may constrain how learners engage with formal scientific knowledge (Vosniadou, 2019). This view is also present in studies that highlight misconceptions in everyday understandings of MLPAs.

### Misconceptions perspective

A number of studies center on misconceptions that emerge from theories that novices have about MLPAs and how novices overestimate the abilities of AI/ML technologies. For instance, arguing that interacting with black-boxed MLPAs may lead to children's development of inaccurate or simplified mental models that can be difficult to overcome (Hitron, 2019). Long and Magerko (2020) claim that misconceptions occur when humans attribute more intelligence to systems than what they have, when they attribute little complexity to systems due to their lack of transparency, and when the systems misguide users on how they work. A common misconception highlighted in several studies is that youths assume that computers learn like humans, have intuition and common sense sometimes being influenced by representations of AI in popular science fiction (Sulmont et al., 2019; Marx, 2022). Another commonly reported misconception is that ML is automated and does not require a lot of human decision making, with ML systems being able to learn by themselves (Sulmont et al., 2019; Große-Bölting & Mühling, 2020). However, previous research also argues that novices may also hold the opposite misconception: that humans decide on all the specifications and create all the rules that ML systems use (Große-Bölting & Mühling, 2020). A third misconception from the literature is that people over-estimate the capacity of ML in solving problems (Sulmont et al., 2019).

### Knowledge-in-Pieces Perspective

While these perspectives offer insights into how young people perceive ML, the focus on attributions, theories, and misconceptions also restricts our understanding of how young people's everyday knowledge can be leveraged for the design of learning tools and activities. As diSessa (2018) argues, everyday knowledge is rich and goes beyond "simple characterizations (e.g., as isolated "misconceptions," simple false beliefs)" (p. 68). Since learning involves building on and rearranging existing knowledge, learners cannot reject their everyday understandings, instead, they can build from them to develop normative understandings. In the learning sciences (LS), diSessa and colleagues (Smith et al., 1994; diSessa, 1993) proposed an alternative approach to studying everyday knowledge that rejects the notion that youths have coherent theories or hypotheses and that decenters misconceptions to argue for a knowledge-in-pieces perspective where youths' everyday knowledge is framed as productive for learning scientific content. Using the knowledge-in-pieces (KiP) perspective we aim to study youths' everyday understandings of ML. The KiP perspective was developed from research on young people's naive understanding of physics, yet in LS it has been used to analyze naive intuitive understandings of phenomena across different domains including computer science (Chao et al., 2018).

diSessa (1993) argued that everyday intuitive knowledge about physics could be organized as a collection of pieces of knowledge or phenomenological primitives (p-prims). These pieces enable people to explain and predict scientific phenomena intuitively and in a way that is consistent with their lived experiences. Yet, these are contextually activated; novices may have inconsistent understandings developed across different contexts. diSessa (2013), recognizing that some understandings may not align with formal scientific knowledge, proposed KiP as a productive take on everyday knowledge. Knowledge pieces are part of a "rich naïve cognitive ecology [that] constitutes a generative pool of resources" (p. 43) that can be productive for learners to attain formal scientific knowledge. Here, learners develop a repertoire of ideas adding understandings from instruction to those from their lived experiences while grappling with multiple and sometimes conflicting ideas they may have about scientific phenomena (Linn, 2006). KiP aims to move away from deficit perspectives on everyday knowledge that highlight misconceptions or wrong theories and hypotheses. Within the KiP approach, "misconceptions" are viewed as having their roots in productive and effective knowledge (Smith et al., 1994). As such, learners' prior conceptions serve as resources for cognitive growth through the reorganization and refinement of ideas. Taking a KiP perspective may be productive to identify youths' everyday understandings of ML and how these could be leveraged for learning how "the ML pipeline"—an abstract model used in instruction to explain how MLPAs function by representing the relationship between data processing operations through which raw data is transformed into features used by a model (Fiebrink, 2019; Schapire, 2014)— works.

In our study, we investigated youths' everyday understanding of ML by presenting teenagers with MLPAs —including commercial products (e.g., TikTok filter, a voice assistant) and e-textile wearables (e.g., a sports cap that reacts to cheering, smart plushies)— and asking them to explain how these worked. We analyzed youths' discourse to identify knowledge components and later compared these with formal models of the "ML pipeline" used in instruction. We address the following research question: What everyday knowledge pieces emerge in youths' explanations of how MLPAs function? With KiP we aimed to investigate pieces of knowledge youths can build on as they learn more about ML. This is very different from the attributions perspective that does not account for *how* MLPAs work, the theory perspective where understandings are seen coherent or the misconceptions approach that focuses solely on inaccurate conceptions.

## Methods

### Participants & workshop design

We conducted an in-person two-hour-long workshop in Spring 2022 at a science center located in the Northeastern United States with 19 teens (ages 14-16) that had demonstrated an interest in STEM by participating in an out of school program designed to deepen participation for historically excluded communities. Consenting participants included 11 teens that had taken part in the science center program for at least one year and thus already knew each other. Four participants self-identified as female, two as non-binary, and five as male. Seven participants identified as Black, two identified as Asian, one as Latinx and one as North African. Nine youths had taken computing courses at school or workshops outside of school, yet none had taken workshops or courses on ML.

The workshop consisted of a series of activities: hello time (10 minutes), warm-up time (15 minutes), how does it work time (25), discussion time (10 minutes), snack time (15 minutes), how does it work time (25), discussion time (10 minutes), and wrap-up time (10 minutes). Hello time, warm-up time and snack time were community building activities for researchers to get to know the participants. During "how does it work time," youths in groups of four received an artifact (see Fig. 1), a video of the artifact being used, and a big piece of paper with a picture of the artifact. It is worth noting that some of the artifacts and videos included failure cases or edge cases where these did not work as expected (e.g., a TikTok filter that did not work for all people). We invited participants to annotate or draw on the big paper to explain how the artifact worked. While participants completed the task, we projected the following questions: "How do you think this artifact works? What role do you think ML/AI plays in making it work?" Two researchers walked around the room, reminded participants to use the paper to explain their ideas and asked questions such as "What makes you think so? How do you think that works?" In this activity, we adopted the big paper cooperative inquiry technique, where participants are given a large piece of paper to collaboratively elaborate on their ideas by drawing or writing (Woodward et al., 2018). Following, during discussion time participants shared their ideas with other groups.

**Figure 1**
Workshop artifacts and their descriptions (hat anonymized for review).

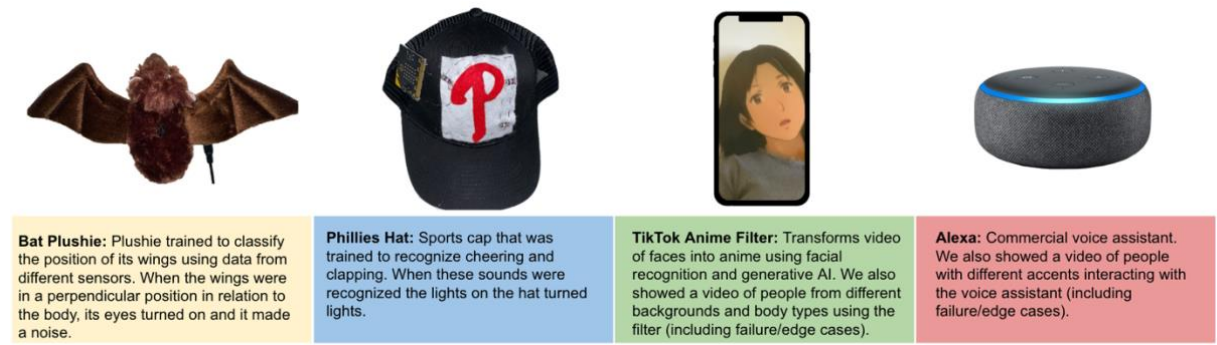

**Bat Plushie:** Plushie trained to classify the position of its wings using data from different sensors. When the wings were in a perpendicular position in relation to the body, its eyes turned on and it made a noise.

**Phillies Hat:** Sports cap that was trained to recognize cheering and clapping. When these sounds were recognized the lights on the hat turned lights.

**TikTok Anime Filter:** Transforms video of faces into anime using facial recognition and generative AI. We also showed a video of people from different backgrounds and body types using the filter (including failure/edge cases).

**Alexa:** Commercial voice assistant. We also showed a video of people with different accents interacting with the voice assistant (including failure/edge cases).

## Data Collection & Analysis

The workshop was led by one of the authors and another researcher acting as facilitators and data collectors. We collected participants' artifacts during both "how does it work time" workshop activities and recorded the discussion youths had during and after the activities. We used an AI/ML transcription tool to process audio recordings and reviewed the transcripts to ensure their accuracy. Following, we identified statements during which youths articulated ideas about the functionality of MLPAs. One author independently coded the big papers and statements from the audio data of one big paper group. This data was coded inductively looking at common themes across the data (Braun & Clarke, 2012). The authors then discussed the coding scheme to define themes. From this round of coding seven categories to classify youths' ideas emerged (training data, data features, inputs, outputs, pattern recognition functions, learning algorithms, testing data). For example, the statement "the filter uses a recognition system for patterns" was coded as pattern recognition. The first author then analyzed the remaining data. Afterwards, we grouped themes into two larger knowledge components related to how MLPAs are trained and how they run. Afterwards, a researcher that had not participated in the project audited the analysis (Lincoln & Guba, 1995), asking questions and identifying points of disagreement. All disagreements and questions were resolved through discussion and social moderation arriving at consensus on our findings.

## **Findings**

We observed that while describing how MLPAs work, youths in the study demonstrated understanding of knowledge pieces related to elements and processes that are part of "the ML pipeline" (see Fig 2.). For instance, they talked about training data or examples being used for applications to "learn" and described the features or properties of data they thought were used. However, they did not elaborate on learning algorithms and did not use the words "model" or "prediction". Instead of talking about models, they mentioned pattern recognition functions that emerge from the training data. In terms of MLPAs running, they claimed that these recognize patterns in input data and depending on the patterns provide different outputs. Our findings show that youths may have pieces of understanding of the ML pipeline based on their observations and everyday experiences with technology. This everyday understanding may be productive for designers and instructors to scaffold youths in building abstract models of how MLPAs work. We group youths' ideas into two knowledge pieces, present their context of emergence and characterize them.

**Figure 2**
Fiebrink's (2019) representation of the ML pipeline as used in instruction (left) and representation of youths' understanding of the ML pipeline (right)

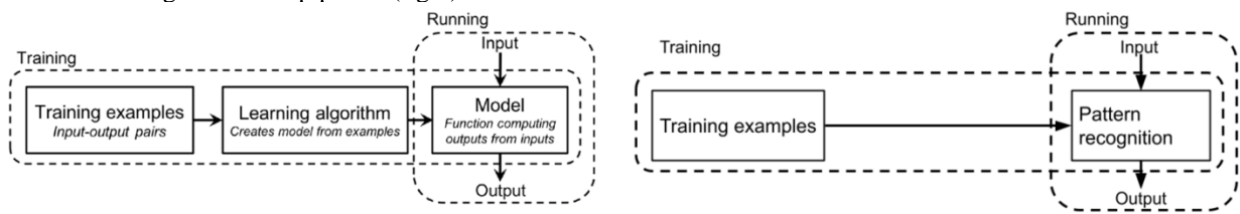

Knowledge piece 1: ML applications learn using training examples.

Youths talked about training data or examples being used for MLPAs to "learn" and described the features or properties of data they thought were used. For instance, Rafik argued that the bat plushie had been trained with data from a gyroscope: "like it was trained against the rotation to the positions [up and down] and so learned eventually what constitutes up and down." He described training examples as input-output pairs. In a similar way, Dwayne expressed that the Phillies hat [team name anonymized for review] used examples in a data set, "basically examples of a cheering sound" implicitly talking about labeled data. Other descriptions of training data were vague with Trisha saying that they "probably have like a database sound to recognize the sound that it is" or Jamar saying that "you put in a Phillies man cheering." Jamar described his understanding in two different ways depending on the context. He insisted that the model included "a programmed example of a crowd cheering and clapping at a Phillies game" but changed his ideas when Armando asked, "What about a Padres game or a non-Phillies game?" This led Jamar to reconsider his explanation to say, "it doesn't matter, it is cheering and clapping, [...] AI recognizes cheering and clapping based on the examples, there will be a pattern when you have examples." Jamar presented fragmented views of what training data could be at first arguing for context specific data (from a Phillies game) and later proposing a more general view.

Another theme that youths discussed were different data features or properties that could be used by MLPAs. Rafik argued that to be able to explain how the bat plushie worked they had "to figure out like what the specific types of data it uses" referring to different sensor data that could be used in the application. With the Alexa and the Phillies' hat that used audio data, youths wondered if the applications recognized sound based on frequency or volume. Carter claimed that "it's all about different parts frequency and volume, it's all part of the same sort of structure with the pattern." Keesha proposed "it depends on how loud it is or maybe it is like listening to diction, would that be the word diction?," while Tia argued it used "vibrations or sound frequency" to find patterns. Similarly, with the TikTok filter, Armando explained that it used specific features such as eyes position and distance and pigments to determine a person's facial structure. These explanations show that youths in the workshop were aware that data may have different features that are used by MLPAs.

It is important to note that youths did not voice ideas related to learning algorithms. On one occasion, Rafik mentioned that data examples were used to train "the neural network or whatever" but provided no explanation of his understanding of neural networks. Testing ML models and using testing data were rarely discussed by youth. Tavon, when explaining why the TikTok filter did not work on everyone's face said, "probably they only had a few people to test it." He also argued that the bat was first trained and then tested to see if it correctly identified the position of the wings.

These findings show that youths had some understanding of the complexity of ML systems, recognizing for instance that different data features could be used to train a model or attributing model behavior to its training data. This challenges findings from research on misunderstandings that claims children ascribe little complexity to ML. At the same time, in contrast to earlier research that claims youths either think ML learns by itself or humans create all the rules, youths voiced a nuanced understanding of MLPAs: being designed by humans who provide data for the "system" to learn. The idea that MLPAs learn from training examples is potentially very productive for learners to further explore, learn and discover how different learning algorithms and MLPAs work. At the same time this knowledge piece is limited in that it does not account for learning algorithms which determine how ML models are built.

## Knowledge piece 2: Applications recognize patterns in input data and depending on the patterns provide different outputs.

While youths did not talk about models, they described that MLPAs recognized patterns derived from training data and that when the applications run data inputs are used in pattern recognition procedures that give outputs (see Fig. 3A). These pattern recognition functions are an incipient idea of a model. Jamar explained that once data is "inside the computer it is going to analyze it and it's going to recognize the pattern." He said that patterns are used to classify "things." Rafik argued that the TikTok filter "probably uses the pattern of like faces [...] these basic facial structures that can be interpreted and used". He further explained that "the pattern can sometimes break and so when the pattern breaks when it gets like [some input] it wasn't expecting." Similarly, Dwayne and Armando explained that when an application such as the Phillies cap is running, the input data is analyzed for patterns based on the training samples. The idea of pattern recognition functions is a good example of how fragmented knowledge can be productive for making sense of ML in youths' everyday lives. Indeed, this everyday understanding of "pattern recognition functions" may be productive for understanding models when building applications or during formal instruction.

**Figure 3**
Trisha and Tia's Big paper explanations of the TikTok filter (left) and Alexa (right).

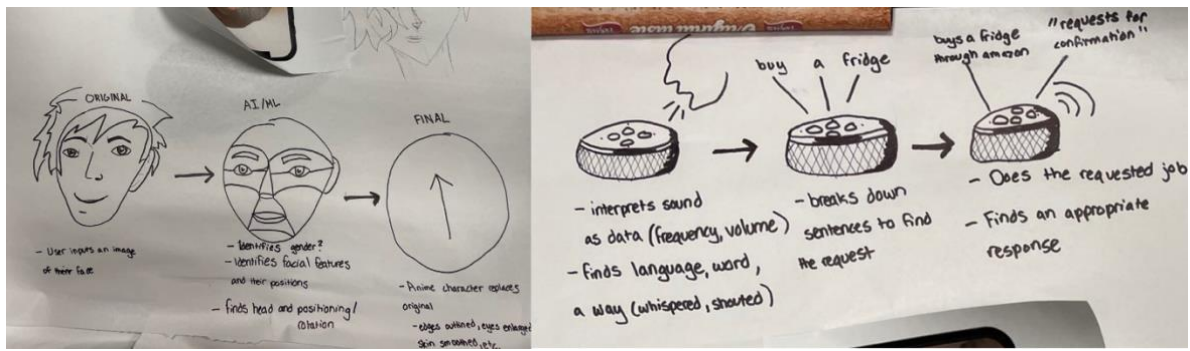

Youths recognized that data was used as an input of the MLPAs, and the outputs of applications depended on these inputs. They discussed what kinds of data could serve as inputs and voiced ideas of how data was input. For instance, Tavon said there was a sensor on the right wing of the bat that was used to know its position, "there are different ways of calculating it [the position of the wings], it knows what position the wing is normally. This is the sensor that calculates the positioning of the wing." Keesha and Tia experimented with different sounds as inputs to see if the Phillies hat reacted to volume, vibrations, or frequency. Tia noticed that where the sound input came from mattered, saying: "it wasn't really lighting up until we made the sound like right under the hat". In terms of outputs, Tia and Armando observed that the lights and sounds from the Phillies hat depended on what sounds it recognized. Rafik and Tavon discussed how the position of the wings of the bat triggered the eyeballs to light up and the bat to make a noise. "When the position of the wings is like this [perpendicular to the body of the bat] the lights turn on," Tavon said. Jamar identified that Alexa's outputs included carrying out the instructions given such as "turning on a light" or "looking for something" as well as providing advertisements related to what people said. While youths did not provide coherent ideas about what specific data inputs and outputs were used by the applications, often disagreeing, their understanding of inputs and outputs may be productive for making sense of the ML pipeline in the future.

Talking about inputs and outputs, youths also discussed some of the limitations of MLPAs and their potential harmful biases. Trisha and Keesha had a long conversation about the TikTok filter and how it only worked on some faces. Keesha explained "is not made for people of color but it definitely is made for light skinned people," Trisha agreed, "it is not made for Black hair." She shared her personal experience explaining she had used a similar filter while wearing an afro and it gave her "a very big forehead." Dwayne wondered what would happen "if the input is someone's face [that] didn't have an eye or something, then the filter might not know what to do," explaining that filters are designed to fit "like the norm." Keesha disagreed saying that filter was not designed for the norm but rather "for an input that fits more of the beauty standard, so we saw that it didn't fit with people who have more weight or people of color." She described the filter as being "Eurocentric." Armando further elaborated, arguing that the problem was not in the input data but in the data used to train the application. "It doesn't have the right training data," he said, making the connection that performance of models depends on the training used. Here youths demonstrated complex understandings of the limitations of MLPAs, for instance of how outputs depend on not only input data but also on the training data used to create the "pattern function." While they did not talk about algorithmic bias, data set diversity or model fitting, their everyday understandings may be productive to make sense of these topics.

## Discussion

We investigated teenagers' everyday understanding of how MLPAs work. Using KiP to analyze youths' ideas, we observed two emerging knowledge pieces: (1) that ML applications learn from training examples and (2) that applications recognize patterns in input data and, depending on the patterns, provide different outputs. Beyond attributions of intelligence, we focused on everyday understanding of *how* these systems work. In contrast to previous research on misunderstandings that highlighted young people ascribing little complexity to ML systems (Long & Magerko, 2020; Sulmont et al., 2019), we saw youths draw inferences that demonstrate some understanding of the complexity of ML particularly on how outputs depend not only on input data but also on training data. Similarly, youths also presented nuanced ideas about how MLPAs are designed by humans who provide data for the "system" to learn. These observations challenge earlier research on misconceptions and theories that describe youths as thinking that MLPAs learn by themselves or that humans have absolute control over how these applications work (Sulmont et al., 2019; Große-Bölting & Mühling, 2020). Whereas youths' everyday understandings do not fully align with formal or normative basic conceptual models of ML, our framing revealed "pieces of knowledge" that closely resemble what instructors may want novices to understand (see Fig.

2). In the following sections, we briefly discuss what these findings contribute to conceptual understanding, further research, and the design of learning activities.

In terms of conceptual understanding, with KiP we aimed to investigate pieces of knowledge youths can build on as they learn more about ML. This productive stance towards everyday knowledge contrasts with the theory perspective, where knowledge is perceived as cohesive hypotheses that are categorized as either correct or incorrect, and the misconceptions approach that centers solely on inaccurate ideas. Furthermore, our comparison between youths' everyday understanding and the "ML Pipeline" (see Fig. 2), as used in instruction, illustrates that everyday knowledge closely resembles the pipeline, and it may be used to scaffold students to achieve instructional goals. It is worth noting that different artifacts elicited different ideas in participants, for instance, interacting with simple applications such as the plushy or the hat provided a context for youths to make inferences about different data features. By providing youths with different contexts to think about ML, we observed that their understandings were contextually bound, with youths building on their own experiences as users to explain how the filter or the voice assistant worked. For instance, showing videos with failure/edge cases also provided a context for youths to consider the applications being used in a different context and to voice understandings related to why these did not work. Our analysis is limited in that it is rather broad as we looked at pieces of knowledge in relation to abstract understanding.

Regarding research methods, this analysis can be helpful for future studies to further investigate everyday conceptual understanding of specific ML concepts in more detail. While we used big paper methods, which enabled us to observe youths explain the functionality of MLPAs while engaging with their peers, most KiP studies have commonly used individual clinical interviews to elicit learners' understanding of phenomena. Future studies could use clinical interviews and other cooperative inquiry techniques to investigate how different methods may elicit different understandings from participants. Similar studies to this one could be conducted prior and post interventions to investigate their influence on conceptual development. At the same time similar studies could also be conducted with different age groups to better understand how children make sense of MLPAs at different stages of development.

Finally, in terms of designing learning activities and tools, the KiP approach to everyday knowledge may be particularly helpful to identify ways in which learning activities can build on what youths may already know. For example, whereas youths did not talk about models, their existing ideas about pattern recognition functions may be productive for introducing ML models. The design of learning activities should consider learner lived experiences and invite them to investigate the technologies they use in everyday life. Finally, toolkits that enable youths to create applications that are like those they use every day may build on the learner's everyday knowledge. Our observations suggest that KiP may provide a nuanced perspective on what youths understand about MLPAs in their everyday lives.

## Acknowledgments

With regards to Yilin Liu for support with data collection and Joouen Shim, Bodong Chen, Xiran Zhu, and Noora Noushad for their feedback.